\documentclass[12pt]{iopart}
\pdfoutput=1
\usepackage{iopams}
\expandafter\let\csname equation*\endcsname\relax
\expandafter\let\csname endequation*\endcsname\relax
\usepackage{amsmath,mathtools}
\usepackage{graphicx}
\usepackage[usenames,dvipsnames]{color}
\usepackage{url}
\usepackage{epstopdf}

\renewcommand{\vec}[1]{\boldsymbol{#1}}

\begin{document}

\title[Exploring the correlation between the folding rates of proteins and entanglement]
{Exploring the correlation between the folding rates of proteins and the entanglement of their native states}

\author{Marco Baiesi}
\author{Enzo Orlandini}
\author{Flavio Seno}
\author{Antonio Trovato}

\address{Dipartimento di Fisica e Astronomia ``Galileo Galilei'',
  Universit\`a di Padova, Via Marzolo 8, 35131, Padova, Italy
} 
\address{INFN, Sezione di Padova, Via Marzolo 8, 35131, Padova,
Italy} 

\begin{abstract}

The folding of a protein towards its native state is a rather complicated process. 
However there are empirical evidences that the  folding time  correlates with the contact order, a simple measure of the 
spatial organisation of the native state of the protein. Contact order  is related to the average length of the main chain loops formed
by amino acids which are in contact.
Here we argue that folding kinetics  can be influenced also  by the 
entanglement that loops may undergo within the overall three dimensional protein structure. 
In order to explore such possibility,  we introduce a novel descriptor, which we call ``maximum intrachain contact entanglement''.
 Specifically, we measure the maximum Gaussian entanglement between any looped portion of a protein and any other non-overlapping subchain of the same protein, which is easily computed by discretized line integrals on the coordinates of the $C_{\alpha}$ atoms.
By analyzing experimental data sets of two-state and multistate folders, we show that also the new index is a good predictor of the folding rate. Moreover, being only partially correlated with previous methods, it can be integrated with them to yield more accurate predictions.
\end{abstract}


\submitto{\JPA}


\maketitle
\section{Introduction}\label{sec:intro}
Simple paradigms very often play an invaluable role to help understanding complex systems. 
A well known example is given by protein folding. Protein folding is the physical process by which a protein chain aquires its final three dimensional structure, the native state, that is usually biologically functional, in a reproducible manner. The characteristic time of this process is named folding time. Protein folding  is complex because of the sheer size of protein molecules, the twenty types of constituent amino acids with distinct side chains, and the essential role played by the environment. Nevertheless, it is by now widely accepted that several aspects  of the process driving a sequence of amino-acids to the corresponding native structure can be inferred by simple descriptors of the native-state geometry~\cite{Baker2000,Dokholyan2002}. 
For instance, it has been shown that the folding 
nucleus of a protein, including the residues whose interactions are essential for the folding to the native state, 
can be predicted through simulations of homopolymer models
based on the mere knowledge of the native contact map, that is of the whole list of residue pairs in contact with each other  ~\cite{Micheletti99,Munoz99,Galitzakya99,Alm99}.
Other  evidences of this simplicity are the ability of effective energy scores, derived by a statistical 
analysis of folded protein structures, to
discriminate real native states among set of competing decoys~\cite{Park96,Samudrala98,Buchete06,maxent,Cossio12} 
and the finding that the universe of possible proteins folds can be derived by simple coarse-grained models of polymers, which capture few universal properties typical of 
all amino-acids~\cite{Marsella06,Zhang06,Cossio10}.

 A further evidence of this emerging simplicity is the
  empirical result of Plaxco and coworkers~\cite{Plaxco98,Plaxco00},
  who found a significant correlation between experimental folding
  rates of proteins, e.g. the inverse folding times, and a simple
  descriptor of the native state organisation, such as the
  \emph{contact order}, that is, the average chemical length (in terms
  of the number of amino-acids) of the loop formed between residues
  which are in contact.  This results is somewhat surprising because
  folding inevitably involves states other than the native one and
  these conformations might affect the kinetic process. Despite some
  evidence that this correlation is weak for proteins belonging to the
  all-$\beta$ structural class ~\cite{MarekHoang03}, later studies
  confirmed correlations between folding rates and other descriptors
  of the native state organisation. These descriptors are \emph{long
    range order}~\cite{Gromiha01}, the \emph{number of native
    contacts}~\cite{Makarov02,Makarov03}, the \emph{total contact
    distance}~\cite{Zhou02}, the
  \emph{cliquishness}~\cite{Micheletti03}, the \emph{local secondary
    structure content}~\cite{Gong03} and the \emph{chain crosslinks
    contact order}~\cite{Dixit06}.

The contact order and all its possible variants are descriptors that
focus on the network of pairs of residues that are nearby in space
regardless of the full spatial arrangement of the
  protein conformation. More realistically, one can however think of
non-local descriptors that capture the degree of self-entanglement of
the whole protein backbone, seen as a curve in a three dimensional
space.  An example is the \emph{writhing number}, a measure of how a
curve winds around itself in space~\cite{Fuller_1971}.  In protein
physics, the writhing number was first used in~\cite{Levitt83} to
quantify the amount of self-threading in native state and it was later
extended to perform a systematic classification of existing protein
folds~\cite{Arteca99,Fain2004}.

After the seminal observation that the backbone of a protein may self entangle into physical knots~\cite{Taylor00,Virnau06,Lua06,Sulkowska08,Mallam08,Mallam10,Jamroz14,Lim15}, a growing attention has been devoted to find new, 
topologically inspired, descriptors for quantifying accurately 
the winding of a protein with itself or with other molecules. 
Specific descriptors have been proposed to measure the amount and location of 
mutual entanglement between protein complexes~\cite{Baiesi16,SulkowskaNAR17,Cieplak17,Caraglio17} or to detect specific topological knots, links and lassos within a single chain~\cite{SulkowskaNAR17,SulkowskaPNAS17,SulkowskaSCREP17}.

The aim of this study is to explore the correlation between protein folding rates and a novel
topological descriptor of proteins three dimensional entanglement, which we
name \emph{maximum intrachain contact entanglement}. 
This indicator
is the maximum value of the mutual entanglement measured between any looped portion of a protein and any other non-overlapping subchain extracted from the same protein. As a measure of the mutual entanglement, 
we consider the Gaussian double integral of two oriented curves.  For closed curves this 
measure reduces to  the Gauss linking number, a topological invariant that quantifies how pairs of loops are  (homologically) linked~\cite{rolfsen1976}.
Being quite easy to compute, the Gauss linking number  
has been extensively used in the past  to characterise the mutual entanglement of 
diluted and concentrated solutions of linear polymers~\cite{doi-edwards,orlandini2000,orlandini2004,Panagiotou_et_al_PRE_2013}, to estimate the linking probability and link complexity  of 
pair of loops under geometrical constraints~\cite{orlandini1994,Arsuaga_et_al_2007,Marko2011,Dadamo_et_al_2017} as well as to  identify threadings in dense solutions
of unlinked loops diffusing  in a gel~\cite{Michieletto_et_al_2014}.

By exploring a data set of 48 proteins~\cite{Micheletti03,Dixit06,Ivankov03} for which
the folding time is known experimentally, we compute the linear correlation coefficient between the values of our 
descriptor and the experimental folding rates. 
We show that the maximum intrachain contact entanglement captures aspects that are
different from those highlighted by the contact order, and we describe how the two descriptors can be 
combined to improve the predictions of folding rates.

\section{Methods}\label{sec:met}

\subsection{A topologically inspired descriptor}

It is well known that the Gauss double integral
\begin{equation}
\label{Gauss-int}
{G} \equiv \frac{1}{4 \pi} \oint_{\gamma_1}\oint_{\gamma_2}
\frac{\vec r^{(1)} -\vec r^{(2)}}{\left| \vec r^{(1)} - \vec r^{(2)}\right|^3} 
\cdot (d \vec r^{(1)} \times d \vec r^{(2)})
\end{equation} 
between two closed curves $\gamma_1$ and $\gamma_2$ in $\mathbb{R}^3$ gives an integer number, 
known as the linking number, whose value is a topological invariant.
A nice feature of this measure, however, is that it provides a meaningful assessment of the 
mutual entanglement also if either one or both curves are open~\cite{doi-edwards,Panagiotou_et_al_PRE_2013,Baiesi16}.

\begin{figure*}[!tb] 
\begin{flushright}
\includegraphics[width=10cm]{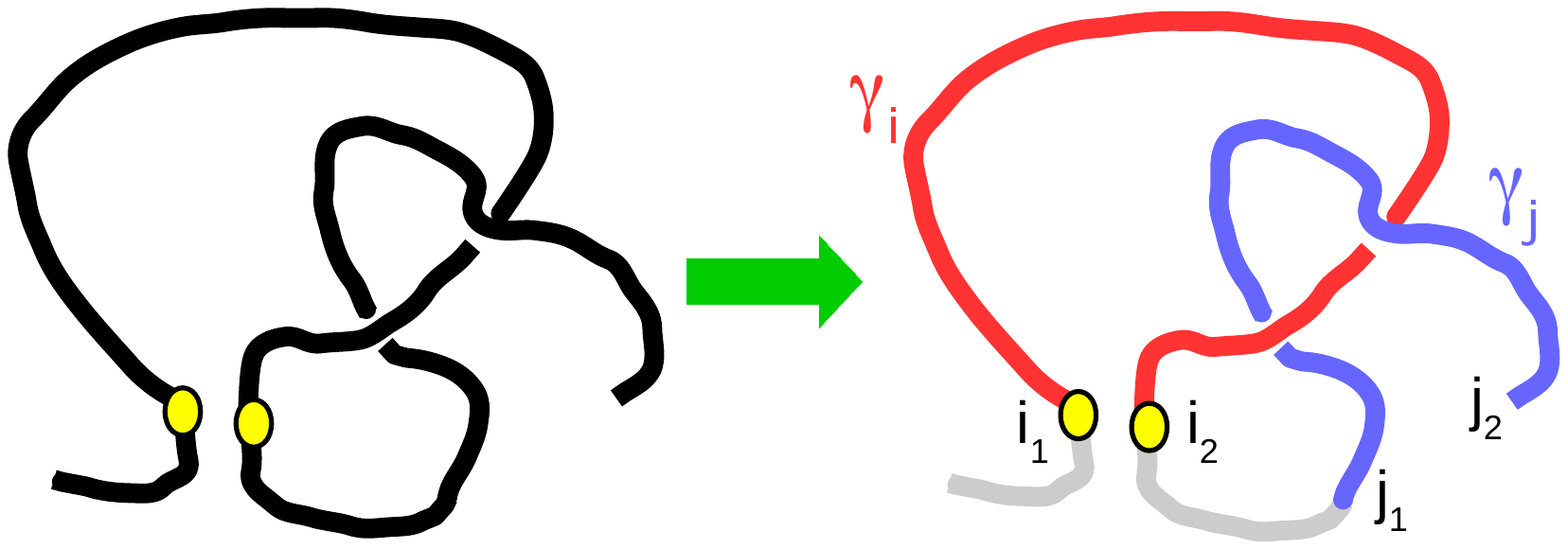}
\end{flushright}
\caption{Sketch of the procedure we use for computing $G'_{ij}$. First, a looped portion $\gamma_i$ ($i_1 \to i_2$, red) is identified when $C_\alpha$ coordinates of amino acids $i_1$ and $i_2$ (yellow ovals) are closer than $d=9\AA$. Then, the double sum (\ref{Gij}) is computed for any other portion $\gamma_j$ ($j_1 \to j_2$, blue)  preceding or following  $\gamma_i$.
}
\label{fig:sk} 

\includegraphics[width=5.3cm]{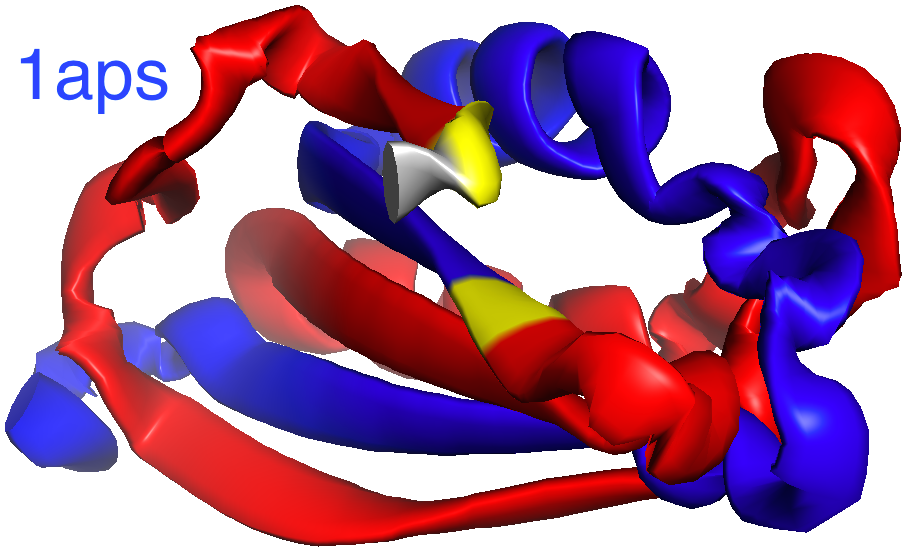}
\includegraphics[width=5.0cm]{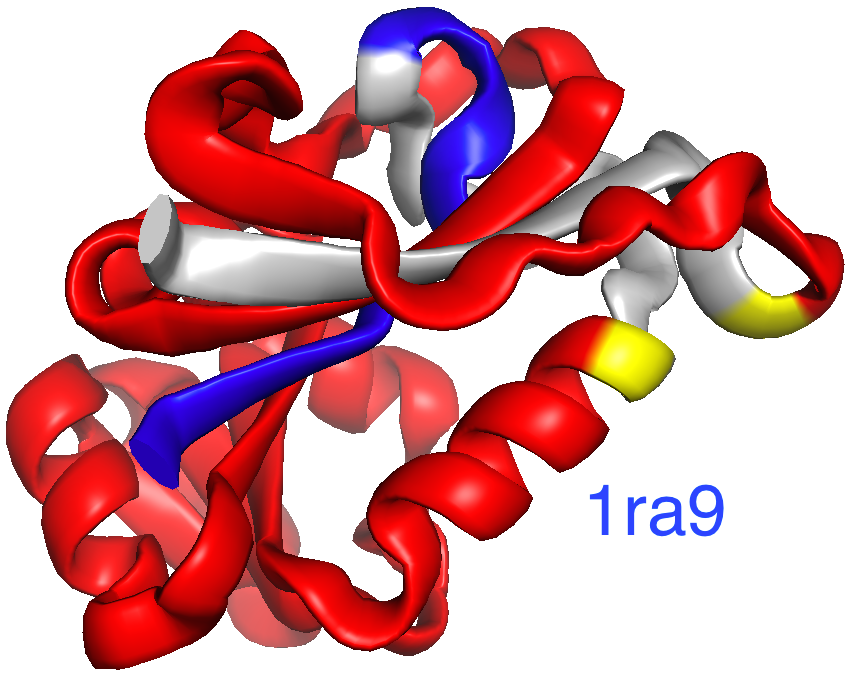}
\includegraphics[width=4.7cm]{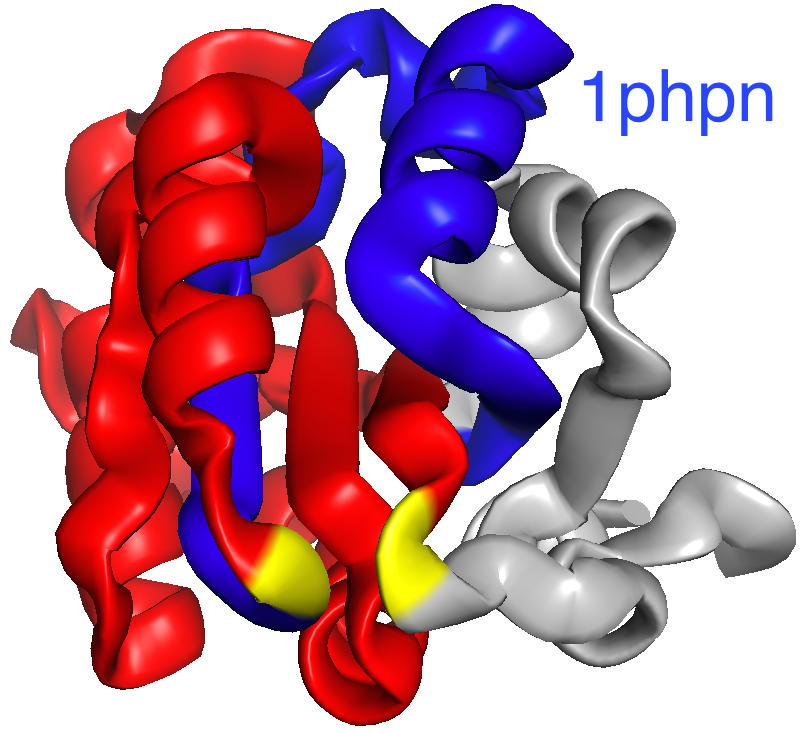}
\caption{Following the color code of figure~\ref{fig:sk}, examples of protein native structures in which we identified the subchains $\gamma_i$ ($i_1 \to i_2$, yellow-red-yellow) and $\gamma_j$ ($j_1 \to j_2$, blue) yielding the maximum intrachain contact entanglement.
}
\label{fig:ex} 
\end{figure*}

Our strategy here is to consider as $\gamma_1$ and $\gamma_2$ any pair $(\gamma_i,\gamma_j)$ 
of non-overlapping subchains extracted from the same protein backbone
and to compute their Gaussian entanglement ${G}_{ij}$.
Since the backbone of a native protein structure  with $N$ residues can be described as a 
discrete chain of monomers ($i=1,\ldots, N$) placed at the positions  $\vec{r}_i$ of 
the C$_\alpha$ atoms, it is natural to define 
the average  positions
\begin{equation}
\vec R_i \equiv \frac 1 2 ( \vec r_i + \vec r_{i+1} ) \,,
\end{equation}
and the bond vectors
\begin{equation}
\vec {d R}_i =  \vec r_{i+1} - \vec r_{i}.
\end{equation}
Hence, for a Given  subchain $\gamma_i$ with monomers from index $i_1$ to $i_2$, and 
a subchain $\gamma_j$ with monomers from index $j_1$ to $j_2$ such that $\gamma_j \cap \gamma_i =\emptyset$,  their Gaussian entanglement is given by
\begin{equation}
\label{Gij}
{G'}_{i j} \equiv \frac{1}{4 \pi} 
\sum_{i=i_1}^{i_2-1} \sum_{j=j_1}^{j_2-1} 
\frac{\vec R_i - \vec R_j}{
\left|\vec R_i - \vec R_j\right|^3} \cdot ( \vec{d R_i} \times \vec{d  R_j})
\end{equation} 
where the prime in $G'$ highlights the fact that the measure is for open chains.  
Note that, unlike in our previous study~\cite{Baiesi16} where the entanglement was estimated between two different 
protein backbones, here the pairs of subchains $(\gamma_i,\gamma_j)$ are extracted from the 
same protein backbone. 

The definition (\ref{Gij}) is rather generic and can be applied to any
pair $(\gamma_i,\gamma_j)$ of not overlapping portions of the protein
backbone. Here we specialize the analysis to the subset of
$(\gamma_i,\gamma_j)$ where the subchain $\gamma_i$ has its first
($i_1$) and last ($i_2$) residues forming a {\em contact} ($i_1\div
i_2$), i.e.~when $|\vec r_{i_1} - \vec r_{i_2}| < d$, with $d=9
\AA$. With this restriction $\gamma_i$ is essentially a loop. The same
restriction is not applied to $\gamma_j$, which can either precede or
follow $\gamma_i$ along the protein backbone.

This way of detecting entangled configurations is sketched in
figure~\ref{fig:sk}.  It is similar to that leading to the definition
of ``lassos''~\cite{SulkowskaSCREP17}, although here we do not
restrict the contacts to chemically strong bonds as in cystein pairs.

A preliminary analysis of our data set showed that such entangled
configurations are not rare. Given their topological complexity, it is
reasonable to think that these native states host proteins which might
fold slowly, especially when the mutual entanglement between
$\gamma_i$ and $\gamma_j$ is considerable.

Motivated by the considerations above, we perform a statistical
analysis to test the existence of a negative correlation between
folding rates and a quantitative measure of the intrachain
entanglement present in a protein native structure. We define this
measure to be the largest absolute value of the mutual entanglement
found for all possible pairs $(\gamma_i,\gamma_j)$ having the
lasso structure discussed above:
\begin{equation}
|G'|_c = \max_{[i_1,i_2],[j_1,j_2]}|G'_{ij}|.
\end{equation}
The index ``c'' indicates that this maximum intrachain {\em contact} entanglement
is subject to the loop constraint of having the ends of $\gamma_i$ in contact with each other, $i_1 \div i_2$.

With the same definition of contact between non consecutive residues
we can introduce the absolute contact order (ACO). This is the average chemical distance $|j-i|$ 
between monomers in contact. Supposing that there are $n_c$ of these contacts in the native state of a protein, we have
\begin{equation}
\textrm{ACO} \equiv \frac 1 {n_c} \sum_{i\div j}|j-i|
\end{equation}
The relative contact order (RCO) is simply the ACO divided by the chain length $N$, which is the average of normalized chemical distances $|j-i| / N$ of residues in contact~\cite{Plaxco98}.

\subsection{Data sets}

We use two separate data sets. A first data set for two-state folders includes
single-domain, non-disulfide-bonded proteins that have been reported
to fold via two-state kinetics under at least some conditions
\cite{Grantcharova01}. We use folding rates as reported previously
\cite{Dixit06,Grantcharova01}, see table \ref{tab:2s}.

\begin{table}
\caption{\label{tab:2s} Data set for two-state folders. $N$ is the
  number of C$_\alpha$ atoms with available coordinates used in the
  computation of $|G'|_c$, \textrm{ACO}, \textrm{RCO}.  1bnza refer to
  chain A in the 1bnz protein-DNA complex. 1div.n refer to the
  N-terminal domain of protein 1div.  1hz6a refer to chain A in the
  1hz6 protein complex.  1lmb3 refer to chain 3 in the 1lmb
  protein-DNA complex. 1urna refer to chain A in the 1urn protein-RNA
  complex.}
\begin{indented}
\item[]\begin{tabular}{@{}llllll}
\br
\textrm{PDB code} & \textrm{ln(rate)} & $N$ & $|G'|_c$ & \textrm{ACO} & \textrm{RCO} \\
\br
\text{1afi} & 0.6 & 72 & 0.77 & 22.99 & 0.32 \\
 \text{1aps} & -1.47 & 98 & 1.62 & 34.1 & 0.35 \\
 \text{1aye} & 6.63 & 80 & 0.27 & 13.83 & 0.17 \\
 \text{1bnza} & 6.95 & 64 & 0.27 & 16.39 & 0.26 \\
 \text{1bzp} & 11.12 & 153 & 0.47 & 24.28 & 0.16 \\
 \text{1csp} & 6.54 & 67 & 0.4 & 19.65 & 0.29 \\
 \text{1div.n} & 6.61 & 56 & 0.84 & 13. & 0.23 \\
 \text{1fkb} & 1.38 & 107 & 0.96 & 32.4 & 0.3 \\
 \text{1hrc} & 8.75 & 104 & 0.56 & 23.56 & 0.23 \\
 \text{1hz6a} & 4.1 & 62 & 0.54 & 17.36 & 0.28 \\
 \text{1imq} & 7.28 & 86 & 0.5 & 18.24 & 0.21 \\
 \text{1lmb3} & 11.01 & 80 & 0.3 & 15.14 & 0.19 \\
 \text{1pgb} & 5.66 & 56 & 0.39 & 17.2 & 0.31 \\
 \text{1poh} & 2.69 & 85 & 0.49 & 27.04 & 0.32 \\
 \text{1psf} & 1.17 & 69 & 0.47 & 20.86 & 0.3 \\
 \text{1shf} & 4.54 & 59 & 0.71 & 18.12 & 0.31 \\
 \text{1ten} & 1.06 & 89 & 0.67 & 28.84 & 0.32 \\
 \text{1tit} & 3.48 & 89 & 0.61 & 30.98 & 0.35 \\
 \text{1ubq} & 7.35 & 76 & 0.47 & 21.82 & 0.29 \\
 \text{1urna} & 2.5 & 96 & 1.15 & 28.34 & 0.3 \\
 \text{1wit} & 0.41 & 93 & 0.72 & 33.35 & 0.36 \\
 \text{256b} & 12.2 & 106 & 0.33 & 15.53 & 0.15 \\
 \text{2abd} & 6.56 & 86 & 0.6 & 21.56 & 0.25 \\
 \text{2ci2} & 4.03 & 64 & 0.68 & 20.23 & 0.32 \\
 \text{2pdd} & 9.67 & 41 & 0.3 & 9.04 & 0.22 \\
 \text{2vik} & 7.48 & 126 & 0.86 & 27.02 & 0.21 \\
\br
\end{tabular}
\end{indented}
\end{table}

The second data set, for multistate folders, is summarised in table \ref{tab:ms} and
 includes proteins that exhibit one or more folding intermediates in water, the
entries $34$-$57$ in table 1 from Ref.~\cite{Ivankov03}. We use
folding rates from that table with two exceptions. For 1ra9, we use
the folding rate reported instead in Refs.~\cite{Kamagata04,Ouyang08}.
We then removed 1cbi and 1ifc from the data set, as they are both
homologous to 1opa, thus sharing essentially the same native
structure. We kept 1opa because it has the intermediate rate among the
three.

\begin{table}
\caption{\label{tab:ms} Data set for multistate folders. $N$ is the
  number of C$_\alpha$ atoms with available coordinates used in the
  computation of $|G'|_c$, \textrm{ACO}, \textrm{RCO}.  1phpn and
  1phpc refer to the N-terminal and, respectively, C-terminal domains
  of 1php. 1qopa and 1qopb refer to the chains A and, respectively, B
  of the 1qopa protein complex.}
\begin{indented}
\item[]\begin{tabular}{@{}llllll}
\br
\textrm{PDB code} & \textrm{ln(rate)} & $N$ & $|G'|_c$ & \textrm{ACO} & \textrm{RCO} \\
\br
\text{1a6n} & 1.1 & 151 & 0.48 & 25.71 & 0.17 \\
 \text{1aon} & 0.8 & 155 & 1.35 & 39.1 & 0.25 \\
 \text{1bni} & 2.6 & 108 & 0.6 & 19.32 & 0.18 \\
 \text{1brs} & 3.4 & 87 & 0.43 & 19.8 & 0.23 \\
 \text{1cei} & 5.8 & 85 & 0.38 & 15.18 & 0.18 \\
 \text{1eal} & 1.3 & 127 & 0.29 & 25.58 & 0.2 \\
 \text{1fnf} & 5.5 & 94 & 0.7 & 28.27 & 0.3 \\
 \text{1hng} & 1.8 & 97 & 0.78 & 31.04 & 0.32 \\
 \text{1opa} & 1.4 & 133 & 0.34 & 29.92 & 0.22 \\
 \text{1ra9} & -2.46 & 159 & 1.65 & 40.71 & 0.26 \\
 \text{1sce} & 4.2 & 101 & 0.46 & 23.41 & 0.23 \\
 \text{1tit} & 3.6 & 89 & 0.61 & 30.98 & 0.35 \\
 \text{1ubq} & 5.9 & 76 & 0.47 & 21.82 & 0.29 \\
 \text{2a5e} & 3.5 & 156 & 0.56 & 15.83 & 0.1 \\
 \text{2cro} & 3.7 & 65 & 0.25 & 13.39 & 0.21 \\
 \text{2lzm} & 4.1 & 164 & 0.36 & 16.07 & 0.1 \\
 \text{2rn2} & 0.1 & 155 & 1. & 39.29 & 0.25 \\
 \text{3chy} & 1. & 128 & 0.98 & 18.96 & 0.15 \\
 \text{1phpc} & -3.5 & 219 & 1.23 & 32.51 & 0.15 \\
 \text{1phpn} & 2.3 & 175 & 1.3 & 36.25 & 0.21 \\
 \text{1qopa} & -2.5 & 268 & 1.43 & 41.16 & 0.15 \\
 \text{1qopb} & -6.9 & 392 & 1.43 & 55.09 & 0.14 \\
\br
\end{tabular}
\end{indented}
\end{table}

Note that two proteins, 1tit and 1ubq, belong to both data sets, since
they are multistate folders in water, while switching to two-state
kinetics upon different conditions.

\section{Results}

For each protein structure in the data sets described in section \ref{sec:met},
we consider four
different descriptors: the chain length $N$, the ACO, the RCO and 
the maximum intrachain contact entanglement ($|G'|_c$).

The values of $|G'|_c$ should be compared to the reference value of 1
found for two closed curves that form a standard Hopf link (the same as for
two flat linked rings). We find
$|G'|_c\ge1$ for $9$ out of the overall 46 proteins analyzed in this
work (see tables \ref{tab:2s},\ref{tab:ms}). The largest value in our study is
$|G'|_c=1.65$ for the multistate protein 1ra9.

\subsection{Two-state proteins}

\begin{figure*}[!tb] 
\includegraphics[width=15.5cm]{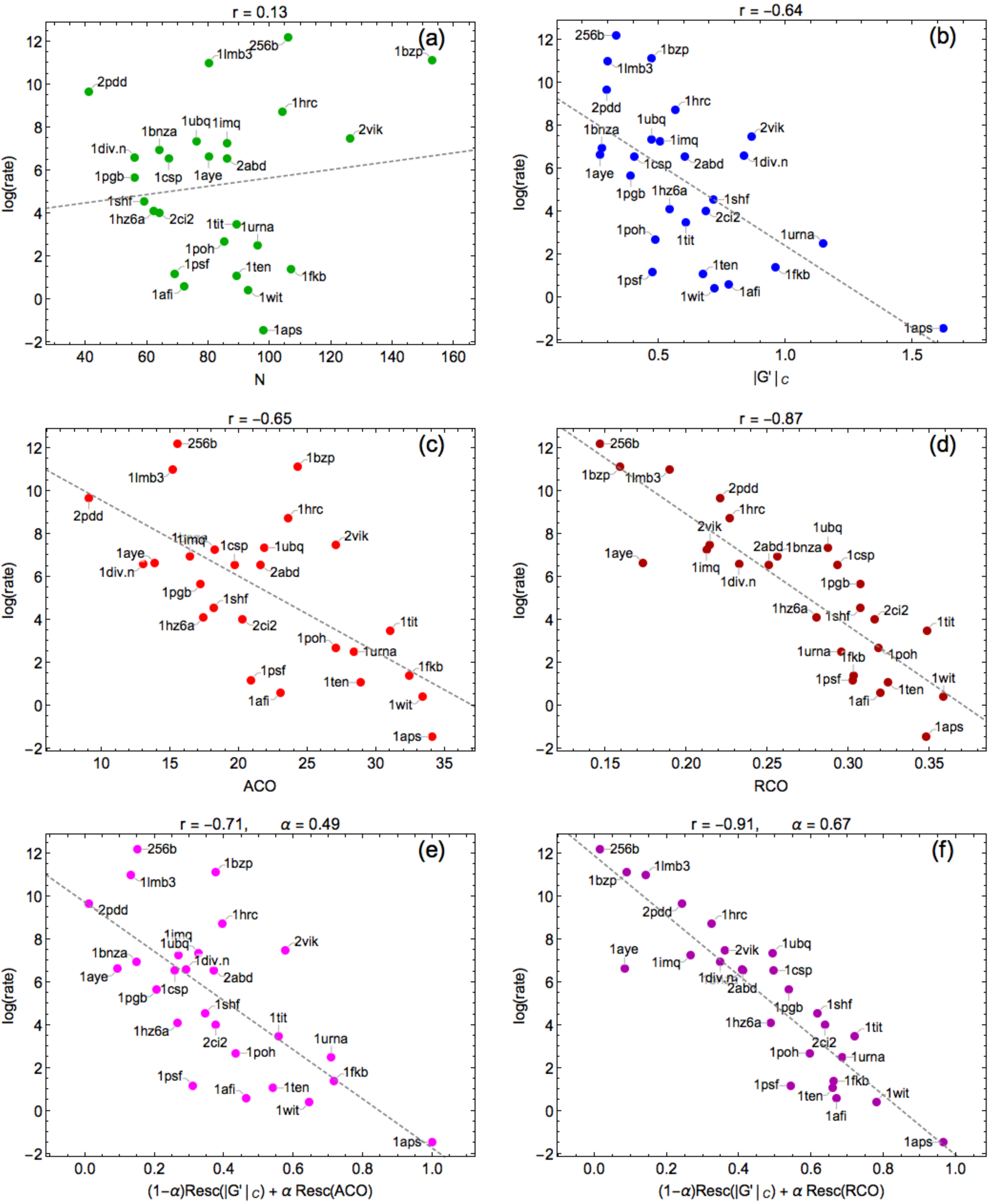}
\caption{For two-state folders,
correlation between the natural logarithm of the folding rate and
(a) chain length, (b) the indicator of entanglement $|G'|_c$ proposed in this work, (c) ACO, 
(d) RCO, (e) a linear combination of rescaled ACO and $|G'|_c$ (see the text), and (f)
a similar linear combination of rescaled RCO and $|G'|_c$. The Pearson
correlation coefficient $r$ of data is specified in all panels.
}
\label{fig:Dixit1} 
\end{figure*}

The linear correlations of the descriptors we consider  with
the natural logarithm of the experimentally measured folding rate are
shown in figure~\ref{fig:Dixit1} for two-state folder together with the 
corresponding Pearson correlation coefficient $r$. As already known
\cite{Dixit06}, chain length is essentially not correlated with the
folding rate ($r=0.13$) for two-state folders (see
figure~\ref{fig:Dixit1}(a)), whereas the best performance (see
figure~\ref{fig:Dixit1}(d)) is achieved by RCO ($r=-0.87$), with negative
correlation implying that slow folders have native structures with
contacting residues that are on average well separated along the
sequence. Although with a lower quality with respect to RCO, the
correlation of ACO with the folding rate of two-state folders is still
significant ($r=-0.65$) and with the proper (negative) slope (see
figure~\ref{fig:Dixit1}(c)).

For the novel topological descriptor that we introduce in this work,
the maximum intrachain contact entanglement $|G'|_c$, the
correlation is essentially as good ($r=-0.64$) as for ACO (see
figure~\ref{fig:Dixit1}(b)).

We next consider how one can  combine linearly  the predicting power of $|G'|_c$ and the contact
order descriptors to achieve correlations with experimental folding
rates that are better than the individual cases. 
To work with homogeneous quantities, acquiring values between $0$ and $1$
in a data set with $k=1,\ldots,N_p$ proteins, we rescale linearly any
descriptor $X_k$ as
\begin{equation}\label{eq:resc}
\mathrm{Resc}\left(X_k\right) = \frac{X_k-X_m}{X_M-X_m}\,,
\end{equation} 
where $X_m = \min_k\left\{X_k\right\}$ and $X_M = \max_k\left\{X_k\right\}$.
The Pearson correlation coefficient is then considered for the linear
combination
\begin{equation}
\label{eq:resc1}
\left(1-\alpha\right)\textrm{Resc}(|G'|_c)
+\alpha\,            \textrm{Resc}(\textrm{ACO})
\end{equation}
as a function of the parameter $\alpha \in [0,1]$ (and similarly for RCO).
Note that  $\alpha=0$ corresponds to consider only  $|G'|_c$ while $\alpha=1$
represents the ACO.

Results for the values of $\alpha$ that yield the higher quality
correlations are shown in figure~\ref{fig:Dixit1}(e) for ACO ($r=-0.71$)
and in figure~\ref{fig:Dixit1}(f) for RCO ($r=-0.91$). In both cases the
performance is increased by combining the contact-order predictor with
the novel entanglement-based predictor. The optimal values of $\alpha$
to be used in the mixing, $\alpha=0.49$ for ACO and
$\alpha=0.67$ for RCO, closer to $0.5$ than to $1$, show  that 
the structural properties captured by $|G'|_c$ are at least in part complementary to
those captured by contact order in the task of predicting folding rates
for two-state folders.

\begin{figure*}[!tb] 
\includegraphics[width=15.5cm]{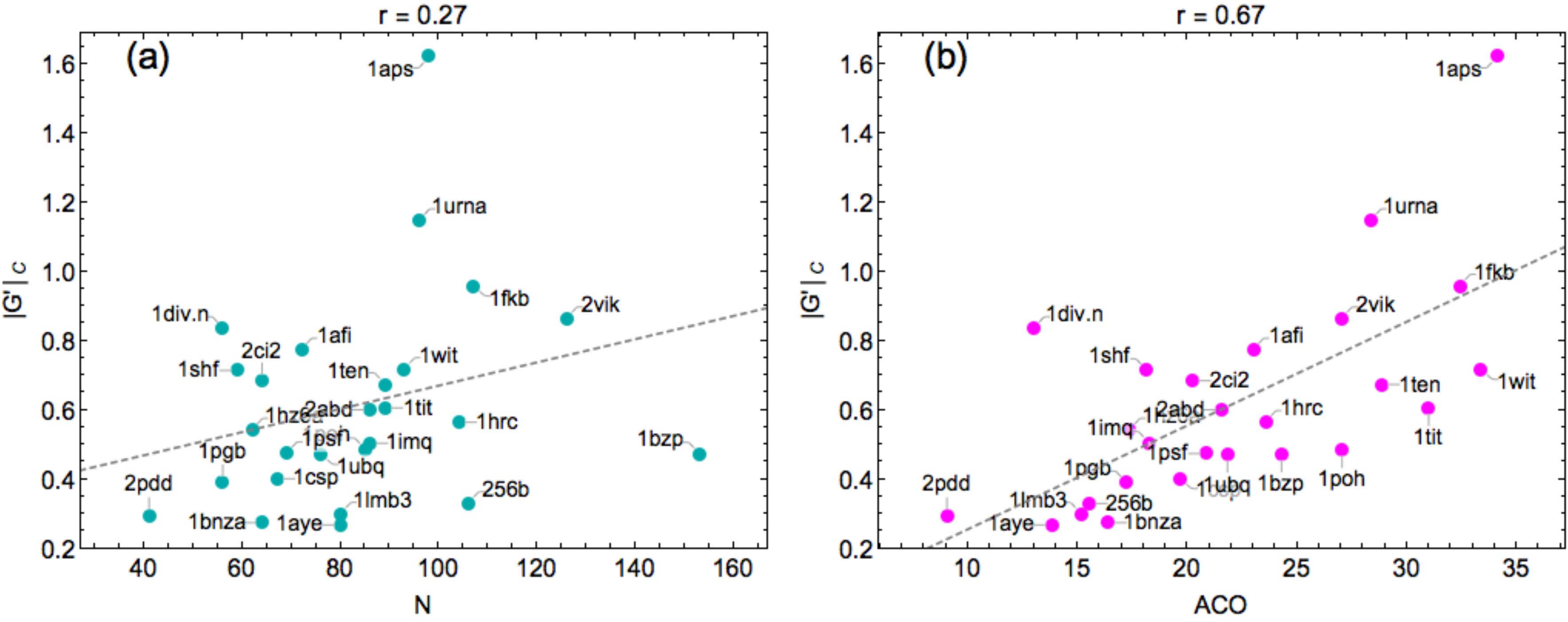}
\caption{For two-state folders, correlation between $|G'|_c$ and (a) chain length and (b) ACO.
}
\label{fig:Dixit2} 
\end{figure*}

Since the increment in the correlation is  related to the
amount of independent information contained in either descriptors,
it important to measure the extent to which  the novel descriptor $|G'|_c$ and 
the other descriptors are mutually correlated.
In figures~\ref{fig:Dixit2}(a),~\ref{fig:Dixit2}(b) we show
the correlation  of $|G'|_c$ with respectively the chain length and ACO.
Structural entanglement, as measured by
$|G'|_c$, is only slightly correlated with chain length ($r=0.27$),
whereas it exhibits a stronger correlation with the ACO ($r=0.67$). Some correlation between $|G'|_c$ and ACO should have been expected, since both quantities correlate well with the experimental results (see Discussion).

\begin{figure*}[!tb] 
\includegraphics[width=15.5cm]{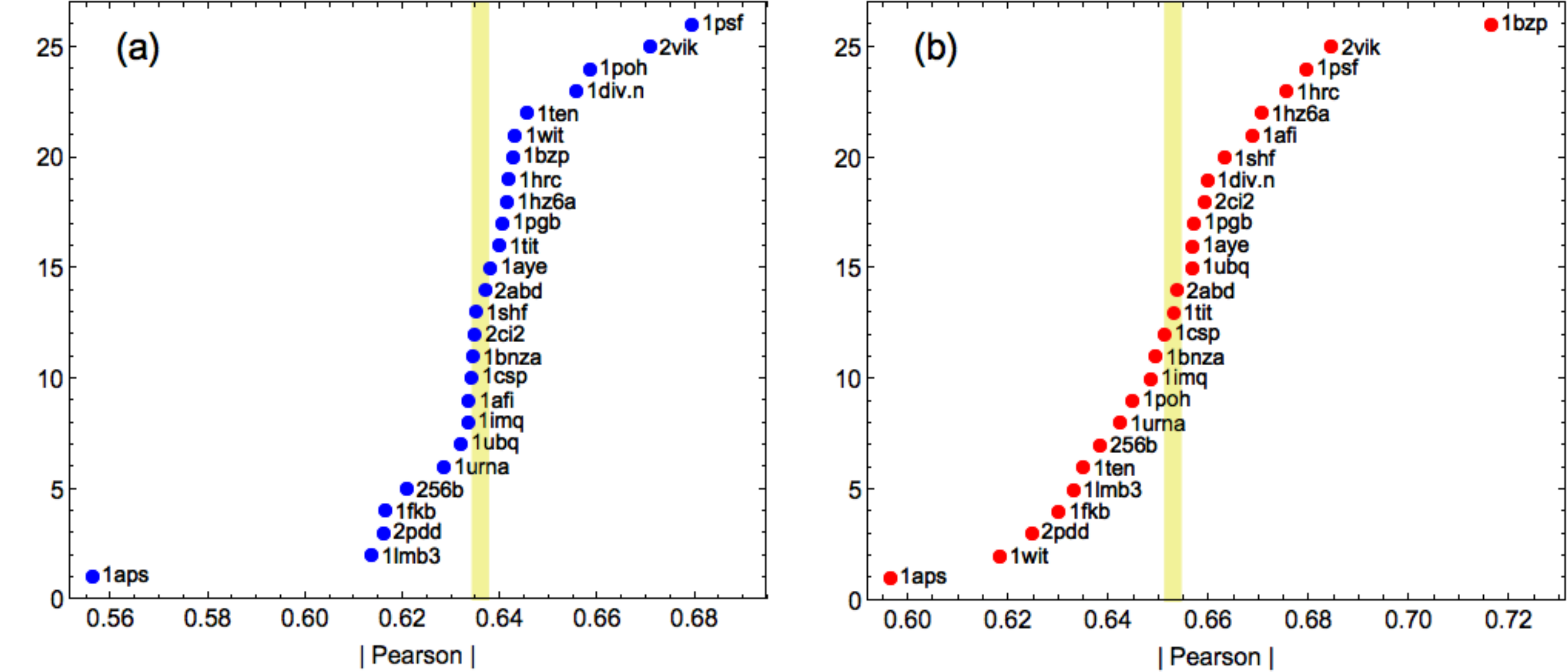}
\caption{For two-state folders: (a) absolute vale of the Pearson correlation coefficient of log-rate vs.~$|G'|_C$ obtained by removing one protein from the database, ranked from the lowest to the highest value. The vertical yellow line indicates the coefficient obtained with the full database. (b) The same for ACO.
In both plots, the protein with the lowest value is the most important for obtaining the Pearson coefficient of the full data set, while the protein with the highest value is the one which spoils mostly the global value.
}
\label{fig:Dixit3} 
\end{figure*}

Finally, we investigate the robustness of the correlation with the
folding rates of two-state folders for two of the considered
descriptors, $|G'|_c$ and ACO. We perform a leave-one-out 
analysis by removing, in turn, each single entry from the data
sets. The Pearson correlation coefficients computed in all such cases
are ranked according to their absolute value in
figure~\ref{fig:Dixit3}(a) for $|G'|_c$ and figure~\ref{fig:Dixit3}(b) for
ACO. As expected for a not so large number of data, the correlation
coefficient can be very sensitive to the removal of single entries
from the data set. In particular, the presence of 1aps is found to be
crucial for the good performances of both descriptors, much more so
for $|G'|_c$, whereas the removal of 1bzp greatly boosts the
performance of ACO.

\subsection{Multistate proteins}

\begin{figure*}[!tb] 
\includegraphics[width=15.5cm]{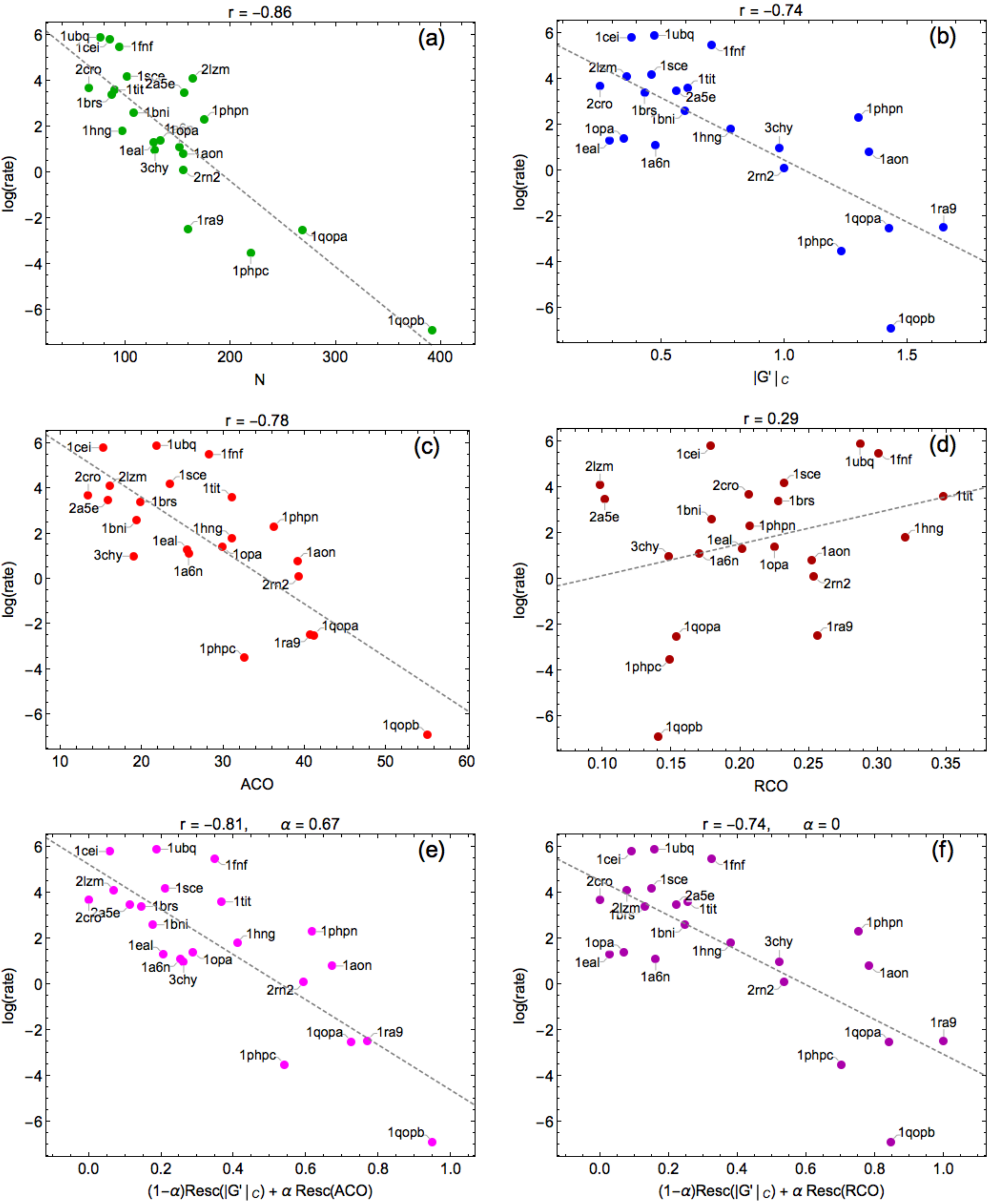}
\caption{As in figure~\ref{fig:Dixit1}, but for multistate folders.
}
\label{fig:Ivankov1} 
\end{figure*}

In analogy with figure~\ref{fig:Dixit1},  for multistate folders
we plot in figure~\ref{fig:Ivankov1} the correlations
between several quantities and the natural logarithm of the experimentally measured folding rate.
As already known~\cite{Weikl08}, the correlation of ACO with folding rate is good
(Pearson correlation coefficient $r=-0.78$) for multistate folders (see figure~\ref{fig:Ivankov1}(c)),
whereas the best performance is 
achieved by chain length ($r=-0.86$, figure~\ref{fig:Ivankov1}(a)).
Contrary to the case of two-state folders, the
correlation of RCO with the folding rate of multistate folders is very poor, 
even reversing its sign ($r=0.29$, see figure~\ref{fig:Ivankov1}(d)).
For $|G'|_c$ the correlation is again almost as good as for ACO ($r=-0.74$, see
figure~\ref{fig:Ivankov1}(b)). 

We next consider how much the linear combination of $|G'|_c$ with
either ACO or RCO can improve the correlation with folding rates of
the contact-order descriptors. In all cases, we rescale linearly the
descriptors $X$ according to (\ref{eq:resc}) and
the Pearson correlation coefficient is then again considered for the linear
combination (\ref{eq:resc1}).

The results for the values of $\alpha$ that yield the higher quality
correlations are shown in figure~\ref{fig:Ivankov1}(e) for ACO ($r=-0.81$)
and in figure~\ref{fig:Ivankov1}(f) for RCO ($r=-0.74$). In both cases the
performance is increased by combining the contact-order predictor with
the novel entanglement-based predictor. The optimal values of $\alpha$
to be used in the linear combination is $\alpha=0.67$ for ACO, confirming
that the structural properties captured by $|G'|_c$ are complementary
to those captured by contact order in the task of predicting folding
rates, also in the case of multistate folders. The linear combination
of $|G'|_c$ with RCO is instead illustrative of the case when one of the
combined predictors ($|G'|_c$) is much more informative than the other, as
evident from the optimal value $\alpha=0$.

\begin{figure*}[!tb] 
\includegraphics[width=15.5cm]{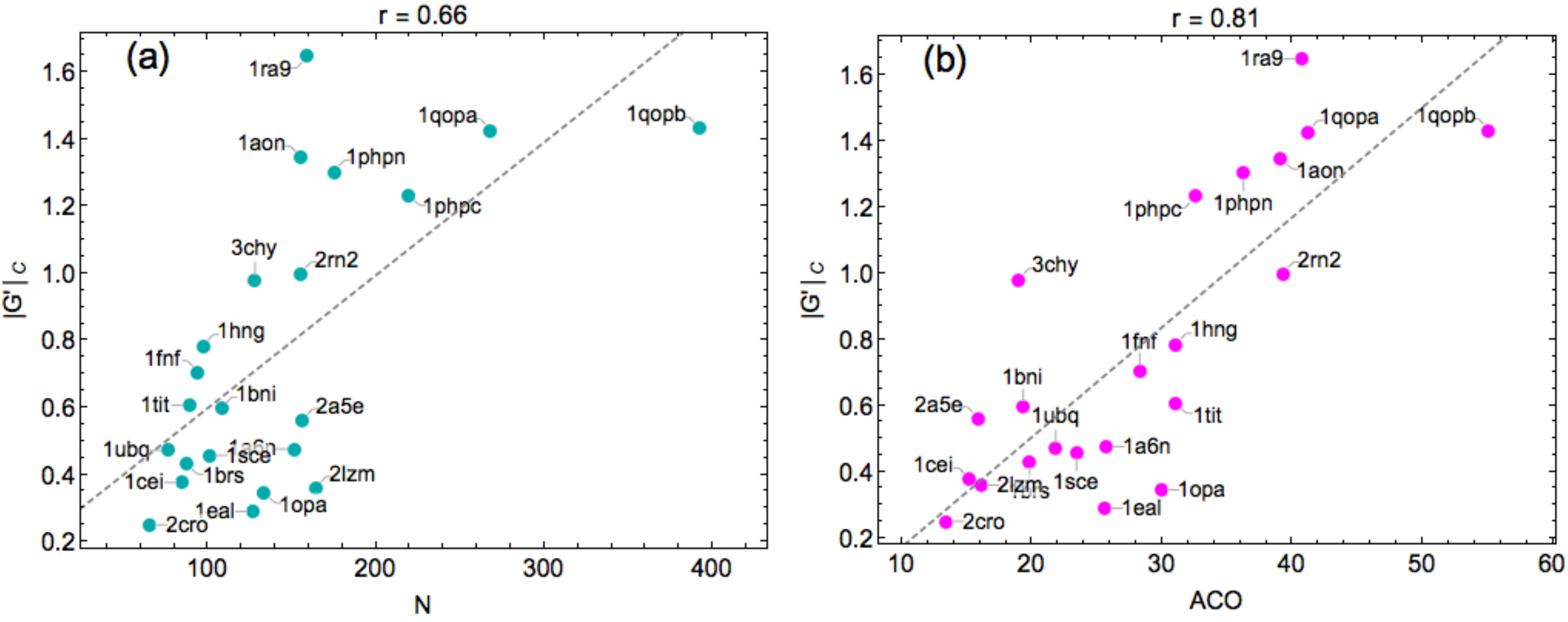}
\caption{For multistate folders, correlation between $|G'|_c$ and (a) chain length and (b) ACO.
}
\label{fig:Ivankov2} 
\end{figure*}

We finally show the correlation of the novel descriptor $|G'|_c$ with
other descriptors, such as chain length (figure~\ref{fig:Ivankov2}(a)) and
ACO (figure~\ref{fig:Ivankov2}(b)). Structural entanglement, as measured by
$|G'|_c$, is significantly correlated with chain length ($r=0.66$),
whereas it exhibits a good correlation with ACO ($r=0.81$).

\begin{figure*}[!tb] 
\includegraphics[width=15.5cm]{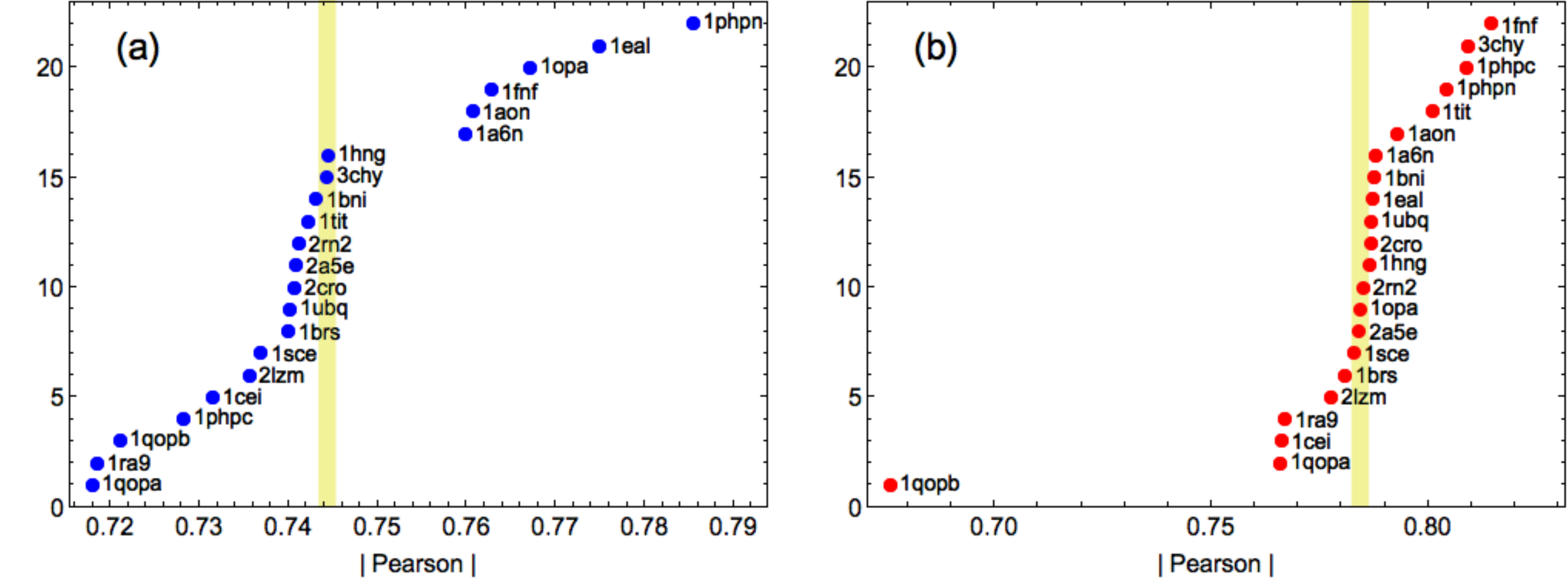}
\caption{For multistate folders: (a) absolute vale of the Pearson correlation coefficient of log-rate vs.~$|G'|_C$ obtained by removing one protein from the database, ranked from the lowest to the highest value. The vertical yellow line indicates the coefficient obtained with the full database. (b) The same for ACO.
}
\label{fig:Ivankov3} 
\end{figure*}

As for the two-state folders, we conclude with a leave-one-out analysis.
Figure~\ref{fig:Ivankov3}(a) suggests that the performance of $|G'|_c$ is robust,
 in this case more than that of ACO, which is very sensible to the presence
of the protein 1qopb in the data set, as shown in figure~\ref{fig:Ivankov3}(b) (the removal
of 1qopb from the data set would cause a drop to $r=0.68$).

\section{Discussion}

Data sets and performance robustness are sensible issues in the context
of folding rate predictions.  Different authors typically considered
different data sets \cite{Weikl08}. Moreover, folding rates may have
been measured for the same protein in different
conditions. Importantly, given the small number of proteins for which
an experimental measure of the folding rate is available, the
performance of different predictors can be very sensitive on the
presence or absence of even single proteins in the data set.

As a matter of fact, several, yet not all, authors considered
separate data sets for two-state and multistate protein folders
\cite{Weikl08}. The folding of two-state proteins to the native state
is a cooperative process characterized by a unique time scale, whose
inverse is the folding rate. Multistate proteins exhibit one or more
intermediate states in the folding process, resulting in multiple
relaxation times being measured. The folding rate of multistate
proteins is associated to the final relaxation to the native state
\cite{Kamagata04}. A given protein may switch from two-state to
multistate folding behaviour upon changing experimental conditions, so
that it can be found in data sets for both categories (see section \ref{sec:met}).
Other authors \cite{Ivankov03,Li04,Naganathan05} considered merged
data sets with both two-state and multistate folders, with the goal of
testing general theories of protein folding that predict how the folding
rate would increase with the chain length, for single-domain proteins.

Our choice here is to keep separate the data sets corresponding to the
two different protein classes. The dependence of folding rate on
simple descriptors, such as chain length or RCO, is in fact very
different in the two classes, see figure~\ref{fig:Dixit1}(a) and (d)
vs.~figure~\ref{fig:Ivankov1}(a) and figure~\ref{fig:Ivankov1}(d). We
are not aware of any effective explanation of this puzzling behaviour.

More generally, the data shown in figure~\ref{fig:Dixit1} and
figure~\ref{fig:Ivankov1} confirm what found in a previous study~\cite{Weikl08}. 
The RCO is a very good predictor for the folding rate
of two-state folders but is very poor for multistate ones, to the
extent of reversing the sign of the correlation coefficient in the
latter case. The converse is true for chain length that is a very
good predictor for the folding rate of multistate folders but essentially does not correlate 
with the folding rate of two-state folders. The ACO is instead a more robust predictor that perform
reasonably well for both data sets and it embodies the ``surprising
simplicity'' that characterizes protein folding \cite{Baker2000}.  The
more topologically complex the network of contacts in the native
structure, as measured in the case of ACO by the average sequence
separation of contacting residue pairs, the longer it takes to fold
to that structure.

Several other descriptors were introduced in the past to capture the
topological complexity of the network of contacts in the native
structures better than ACO. These include long range order, total
contact distance, cliquishness, logCO, number of non local contacts,
and number of geometric contacts
\cite{Gromiha01,Makarov02,Zhou02,Micheletti03,Gong03,Dixit06,Ouyang08}. All
such descriptors are based on the notion of pairwise residue contacts.
The performances of the different descriptors in predicting folding
rates vary somewhat depending on the considered data sets
\cite{Weikl08}. It is fair to state that most of the cited predictors,
including ACO, exhibit overall similar performances.

In this work, in fact, our main focus was not to establish which is
the best predictor of folding rates nor to build such an optimal
algorithm.  We instead introduced a novel descriptor, the maximum
intrachain contact entanglement $|G'|_c$, not directly related to the
contact order. It is rather based on the concept of the mutual
entanglement between two portions of a protein chain that is
inherently associated to contact formation. We then showed that
$|G'|_c$ can be used to predict folding rates with a performance
comparable to the one achieved by ACO, for both data sets of two-state
and multistate folders, as shown in figure~\ref{fig:Dixit1}(b),(c) and
in figure~\ref{fig:Ivankov1}(b),(c).  Figure~\ref{fig:Dixit3} further
shows that ACO performance is slightly more robust for two-state
folders, whereas figure~\ref{fig:Ivankov3} shows that $|G'|_c$
performance is instead more robust for multistate folders.

Our main message is related to the complementary nature of
 the $|G'|_c$ and ACO descriptors in capturing the
topological complexity of protein native structures at two different
levels. Not only the separation along the sequence between pairs of
contacting residues is important, but also the possible entanglement
of other chain portions with the loop connecting two contacting
residues (see figure~\ref{fig:sk}) plays a relevant role.
 Note that the former feature refers
  to the topological complexity of the network of native contacts,
  whereas the latter relates to the topology of the protein chain as a
  curve in the three-dimensional space. The explicit consideration of
  the three-dimensional topological properties of the native structures
 represents one of the main
  novelties of the present paper.

The two descriptors $|G'|_c$ and ACO are indeed correlated for both data sets of
two-state and multistate folders, as shown in
figure~\ref{fig:Dixit2}(b) and in
figure~\ref{fig:Ivankov2}(b). However, a linear combination of
$|G'|_c$ and ACO, after proper rescaling of the two quantities
achieves a better performance than ACO alone, or $|G'|_c$ alone, for
both data sets, as shown in figure~\ref{fig:Dixit2}(e) and in
figure~\ref{fig:Ivankov2}(e). Similarly, a linear combination of
$|G'|_c$ and RCO, after proper rescaling of the two quantities
achieves a better performance than RCO alone for two-state folders, as
shown in figure~\ref{fig:Dixit2}(f).

The definition we chose for the entanglement descriptor requires that
one of the two subchains is looped, and hence that the structure
identified by a high $|G'|_c$ resembles a lasso. This definition is in
line with similar analyses in the literature.  Nevertheless, as Gauss
double integrals do not require the looping condition for any of the
two subchains, more flexible descriptors may be put forward to assess
the degree of entanglement.  General Gauss double integrals thus
constitute a method for future characterisations of the topological
complexity of single proteins, which may find applications
  also in contexts different from the prediction of folding rates
  considered in this work.

As an illustrative example, we conclude our discussion by reporting
that $|G'|_c=1.21$ for the Human single-domain protein K-Ras (167
residues as a single chain in the PDB ID 3GFT). K-Ras fluctuations in
its native ensemble were recently shown by atomistic simulations to
exhibit anomalous non-ergodic kinetics over several decades
\cite{Smith16}. The same behaviour was reported for two larger
multi-domain proteins. The observed kinetics was quantitatively well
described by a continuous time random walk with a heavy-tailed waiting
time distributions \cite{Metzler14}.

The maximum intrachain contact entanglement found for K-Ras is higher
than the one expected for single-domain proteins with similar length
based on a linear interpolation for the data sets considered in this
paper (see figures~\ref{fig:Dixit2}(a),\ref{fig:Ivankov2}(a)).
It may be then appealing to speculate whether the locking of the protein
chain into conformations that are entangled, according to the $|G'|_c$
descriptor, or to similar ones, could play some role
in shaping the non-ergodic kinetics described above. Highly entangled
conformations could in fact explain the presence of deep traps in the
energy landscape where the protein chain would remain stuck for
extended periods of time.
Clearly, such hypothesis needs to be thoroughly validated by further
studies. The generalization of the maximum intrachain contact
entanglement indicator to the case of multi-domain proteins should
also be considered in this respect.

\ack We thank Stu Whittington for having introduced us into the exciting subject of 
topology of random polymers and for having been an inspiring teacher and friend. 
We also thank an anonymous referee for pointing out reference \cite{Smith16}.\\

\end{document}